\documentclass[aps,prb,twocolumn,superscriptaddress]{revtex4-1}
\usepackage[center]{titlesec}
\usepackage{graphicx}

\begin{document}


\title{Ultrafast terahertz-field-driven ionic response in ferroelectric BaTiO$_3$}


\author{F. Chen}
\thanks{These two authors contributed equally to this work}
\affiliation{Department of Electrical Engineering, Stanford University, Stanford, CA 94305}
\affiliation{SIMES Institute for Materials and Energy Sciences, SLAC National Accelerator Laboratory, Menlo Park, CA 94025}
\author{Y. Zhu}
\thanks{These two authors contributed equally to this work}
\affiliation{Advanced Photon Source, Argonne National Laboratory, Argonne, IL 60439}

\author{S. Liu}
\affiliation{The Makineni Theoretical Laboratories, Department of Chemistry, University of Pennsylvania, Philadelphia, PA 19104-6323, USA}
\affiliation{Geophysical Lab, Carnegie Institute for Science, Washington DC 20015, USA}
\author{Y. Qi}
\affiliation{The Makineni Theoretical Laboratories, Department of Chemistry, University of Pennsylvania, Philadelphia, PA 19104-6323, USA}
\author{H.Y. Hwang}
\author{N.C. Brandt}
\author{J. Lu}
\affiliation{Department of Chemistry, Massachusetts Institute of Technology, Cambridge, MA 02139}
\author{F. Quirin}
\affiliation{Faculty of Physics and Center for Nanointegration Duisburg-Essen (CENIDE), University of Duisburg-Essen, Lotharstrasse 1, 47048 Duisburg, Germany}
\author{H. Enquist}
\affiliation{MAX IV Laboratory, Lund University, S-22100 Lund, Sweden}
\author{P. Zalden}
\affiliation{SIMES Institute for Materials and Energy Sciences, SLAC National Accelerator Laboratory, Menlo Park, CA 94025}
\author{T. Hu}
\author{J. Goodfellow}
\affiliation{Department of Materials Science and Engineering, Stanford University, Stanford, CA 94305}
\author{M.-J. Sher}
\affiliation{Department of Materials Science and Engineering, Stanford University, Stanford, CA 94305}
\affiliation{SIMES Institute for Materials and Energy Sciences, SLAC National Accelerator Laboratory, Menlo Park, CA 94025}
\author{M.C. Hoffmann}
\author{D. Zhu}
\author{H. Lemke}
\author{J. Glownia}
\author{M. Chollet}
\affiliation{Linac Coherent Light Source, SLAC National Accelerator Laboratory, Menlo Park, CA 94025}
\author{A. R. Damodaran}
\affiliation{Department of Materials Science and Engineering, University of California Berkeley, Berkeley, CA 94720}
\author{J. Park}
\affiliation{Department of Materials Science and Engineering, University of Wisconsin, Madison, Madison, Wisconsin 53706, USA}
\author{Z. Cai}
\affiliation{Advanced Photon Source, Argonne National Laboratory, Argonne, IL 60439}
\author{I.W. Jung}
\affiliation{Center for Nanoscale Materials, Argonne National Laboratory, Argonne, Illinois 60439, USA}
\author{M.J. Highland}
\affiliation{Materials Science Division, Argonne National Laboratory, Argonne, Illinois 60439, USA}
\author{D.A. Walko}
\author{J. W. Freeland}
\affiliation{Advanced Photon Source, Argonne National Laboratory, Argonne, IL 60439}
\author{P.G. Evans}
\affiliation{Department of Materials Science and Engineering, University of Wisconsin, Madison, Madison, Wisconsin 53706, USA}
\author{A. Vailionis}
\affiliation{Geballe Laboratory for Advanced Materials, Stanford University, Stanford, CA 94305}
\author{J. Larsson}
\affiliation{Department of Physics, Lund University, S-22100 Lund, Sweden}
\author{K.A. Nelson}
\affiliation{Department of Chemistry, Massachusetts Institute of Technology, Cambridge, MA 02139}
\author{A.M. Rappe}
\affiliation{The Makineni Theoretical Laboratories, Department of Chemistry, University of Pennsylvania, Philadelphia, PA 19104-6323, USA}
\author{K. Sokolowski-Tinten}
\affiliation{Faculty of Physics and Center for Nanointegration Duisburg-Essen (CENIDE), University of Duisburg-Essen, Lotharstrasse 1, 47048 Duisburg, Germany.}
\author{L. W. Martin}
\affiliation{Department of Materials Science and Engineering, University of California, Berkeley, Berkeley, CA 94720}
\affiliation{Materials Science Division, Lawrence Berkeley National Laboratory, Berkeley, CA 94720}
\author{H. Wen}
\thanks{Co-corresponding authors: aaronl@stanford.edu and wen@aps.anl.gov}
\affiliation{Advanced Photon Source, Argonne National Laboratory, Argonne, IL 60439}
\author{A.M. Lindenberg}
\thanks{Co-corresponding authors: aaronl@stanford.edu and wen@aps.anl.gov}
\affiliation{SIMES Institute for Materials and Energy Sciences, SLAC National Accelerator Laboratory, Menlo Park, CA 94025}
\affiliation{Department of Materials Science and Engineering, Stanford University, Stanford, CA 94305}
\affiliation{PULSE Institute, SLAC National Accelerator Laboratory, Menlo Park, CA 94025}


\date{\today}

\begin{abstract}
The dynamical processes associated with electric field manipulation of the polarization in a ferroelectric remain largely unknown but fundamentally determine the speed and functionality of ferroelectric materials and devices. Here we apply sub-picosecond duration, single-cycle terahertz pulses as an ultrafast electric field bias to prototypical BaTiO$_3$ ferroelectric thin films with the atomic-scale response probed by femtosecond x-ray scattering techniques. We show that electric fields applied perpendicular to the ferroelectric polarization drive large amplitude displacements of the titanium atoms along the ferroelectric polarization axis, comparable to that of the built-in displacements associated with the intrinsic polarization and incoherent across unit cells.   This effect is associated with a dynamic rotation of the ferroelectric polarization switching on and then off on picosecond timescales. These transient polarization modulations are followed by long-lived vibrational heating effects driven by resonant excitation of the ferroelectric soft mode, as reflected in changes in the c-axis tetragonality.  The ultrafast structural characterization described here enables direct comparison with first-principles-based molecular dynamics simulations, with good agreement obtained.  
\end{abstract}


\maketitle



Ferroelectric materials comprise non-centrosymmetric unit cells with permanent electric dipole moments switchable by electric fields and exhibit strong coupling between polarization, strain, and electronic degrees of freedom. Strong light-matter coupling and photoferroelectric responses associated with these materials have enabled next-generation photovoltaic applications as well as novel optical detection technology spanning the range from visible to far infrared frequencies \cite{Grinberg2013, Choi2009, Chen2014, Yang2010, Young2012}.  In recent years, new possibilities to manipulate the functional properties of ferroelectrics with light have emerged, holding promise both for directing these coupled degrees of freedom and for elucidating their fundamental dynamical properties ~\cite{KorffSchmising2007, Miyamoto2013, Wen2013, Daranciang2012, Kampfrath2013, Chen2015, Kubacka2014, Kundys2010}. In particular, at the heart of ferroelectric-based next generation piezoelectric, electrocaloric, electro-optic, and non-volatile memory devices and sensors lies the dynamics of electric polarization ~\cite{Shin2007, Grigoriev2006, Jiang2012, Grinberg2009, Zenkevich2014}. Although studies of electric-field-driven polarization dynamics have been carried out in the past through the application of short electrical pulses, these are complicated by difficulties coupling electrical pulses through electrode structures on sub-100-picosecond (ps) time-scales, whereas the intrinsic atomic-scale response is orders of magnitude faster. Prior studies ~\cite{Li2004, Grigoriev2006} have shown evidence for sub-nanosecond time-scale dynamics in response to electrical bias fields.  Optical ~\cite{Miyamoto2013, Dougherty1994, Katayama2012} and x-ray studies ~\cite{KorffSchmising2007, Istomin2007, Daranciang2012, Zamponi:2012dz} have captured information about the dynamics of ferroelectrics upon optical excitation whereas the intrinsic atomic-scale response to applied electric fields is largely unexplored.  Whereas prior theoretical work has been able to study the influence of electric-field driven dynamics in ferroelectrics on short time-scales~\cite{Grinberg2009,Qi2009,Shin2007}, direct comparisons between MD simulations and the actual atomic-scale response have not been previously carried out. Here we use single-cycle, sub-ps duration terahertz (THz) pulses as an all-optical means to apply an electric field bias to the prototypical ferroelectric BaTiO$_3$ (BTO), while resolving the time-dependent atomic-scale response $\emph{in-situ}$ using femtosecond x-ray scattering techniques.  In particular, we obtain a direct view of the atomic displacements within the unit cell upon THz excitation, resonant with the soft mode of BTO. Experimental measurements are  compared with state-of-the-art molecular dynamics (MD) simulations, providing a microscopic picture of the atomic-scale response to ultrafast electric field stimulation on ps time-scales. 

\begin{figure}
\includegraphics[width=\columnwidth]{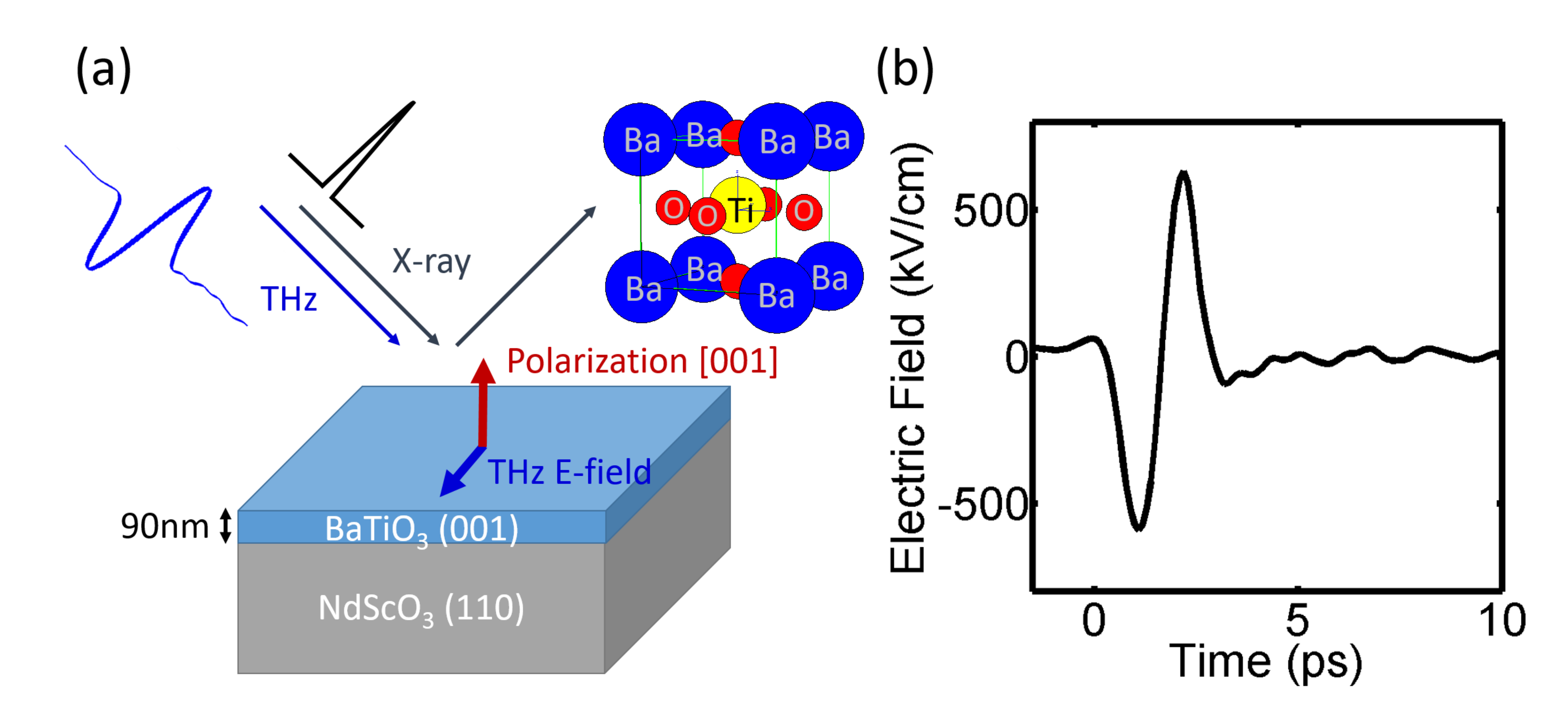}
\caption{Experimental setup, applied THz field, and enhancement electrode structure. (a) Sample schematic showing ferroelectric polarization perpendicular to the film along [001].  The field is applied with light polarization orthogonal to the ferroelectric polarization. The inset shows the pseudocubic unit cell of BTO.  (b)  Applied THz electric field as a function of time, measured by electro-optic sampling at the sample position at LCLS.}
\label{Fig1}
\end{figure}

\begin{figure}
\includegraphics[width=\columnwidth]{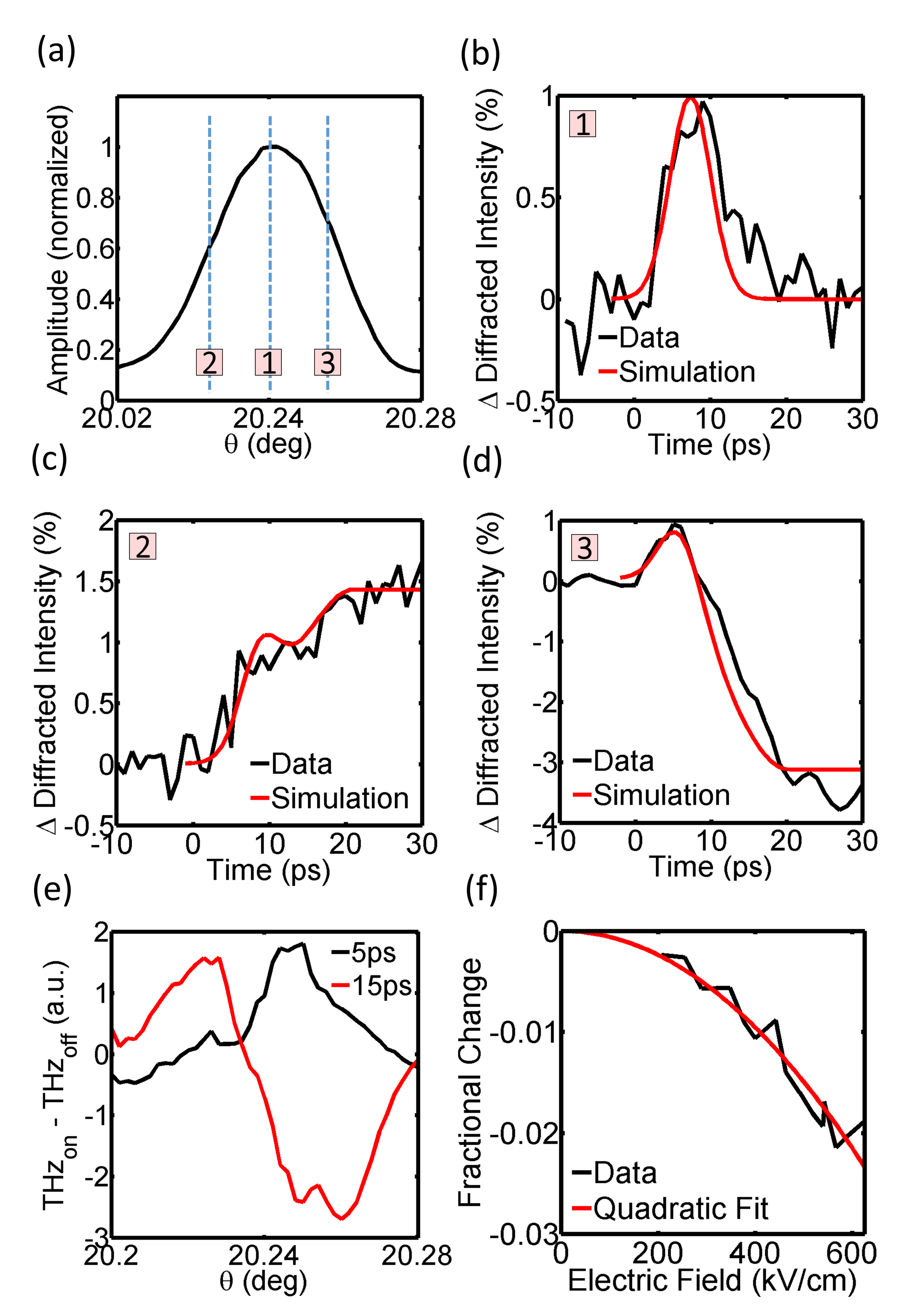}
\caption{Terahertz-pump x-ray probe measurements and field dependence. a) Static BTO (003) rocking curve at room temperature,  with three specific incident angles marked as 1,2 and 3 corresponding to the angles where the time scans (b), (c), and (d) were measured. Red curves are simulations using dynamical diffraction theory based on calculated THz-driven strains and structure factor modulations.  (e) the differential rocking curve at t=5 and 15 ps comparing the rocking curves measured with and without THz excitation.  (f) Dependence of change in diffracted intensity on THz peak field, measured at t=15 ps on the high-angle side of the rocking curve.  Red curve is quadratic fit.  }
\label{Fig2}
\end{figure}

THz pump/x-ray probe measurements with $\approx$200 femtosecond temporal resolution were carried out at the Linac Coherent Light Source (LCLS) at the SLAC National Accelerator Laboratory probing the time-dependent structural response of 90 nm single domain BTO thin films deposited on NdScO$_3$ substrates, with ferroelectric polarization pointing normal to the surface.  Additional measurements on the same sample were carried out at the Advanced Photon Source at Argonne National Laboratory using micro-focused 100 ps x-ray pulses~\cite{Zhu:2016cp} to probe the response at higher fields within split-ring resonator structures (See Supplemental Materials for further details on the experimental setup). Motivated in part by recent theoretical studies~\cite{Qi2009}, the THz field was applied in-plane and perpendicular to the ferroelectric polarization.  Transient THz-induced responses were measured for specific points on the (003) BTO x-ray rocking curve corresponding to the maximum, and half intensity at lower and higher angles of the diffraction peak [Fig. \ref{Fig2}(a,b,c,d)].  At the peak of the rocking curve [Fig. \ref{Fig2}(b)] where one is insensitive to rocking curve shifts, we observe a fast transient increase in the diffracted intensity of approximately 1$\%$, seen also in the measurements on the lower angle and higher angle sides near time-zero [Fig. \ref{Fig2}(c,d)]. This short time increase is associated with an x-ray structure factor increase occurring on a time-scale comparable to the THz pulse duration and lasting of order 10 ps, discussed further below. Following the transient increase observed at all points on the rocking curve in the first 10 ps, we observe a time-dependent increase (decrease) in scattering intensity on the lower angle (higher angle) sides, associated with the development of a long-lived shift of the diffraction peak to lower angles with an onset time of $\approx$15 ps. This is consistent with a homogeneous stress induced by the THz field where the observed time scale is determined by the sample thickness divided by the longitudinal sound velocity (90 nm / 6000 m/s = 15 ps)~\cite{Nicoul2011}.

\begin{figure}[t]
\includegraphics[width=\columnwidth]{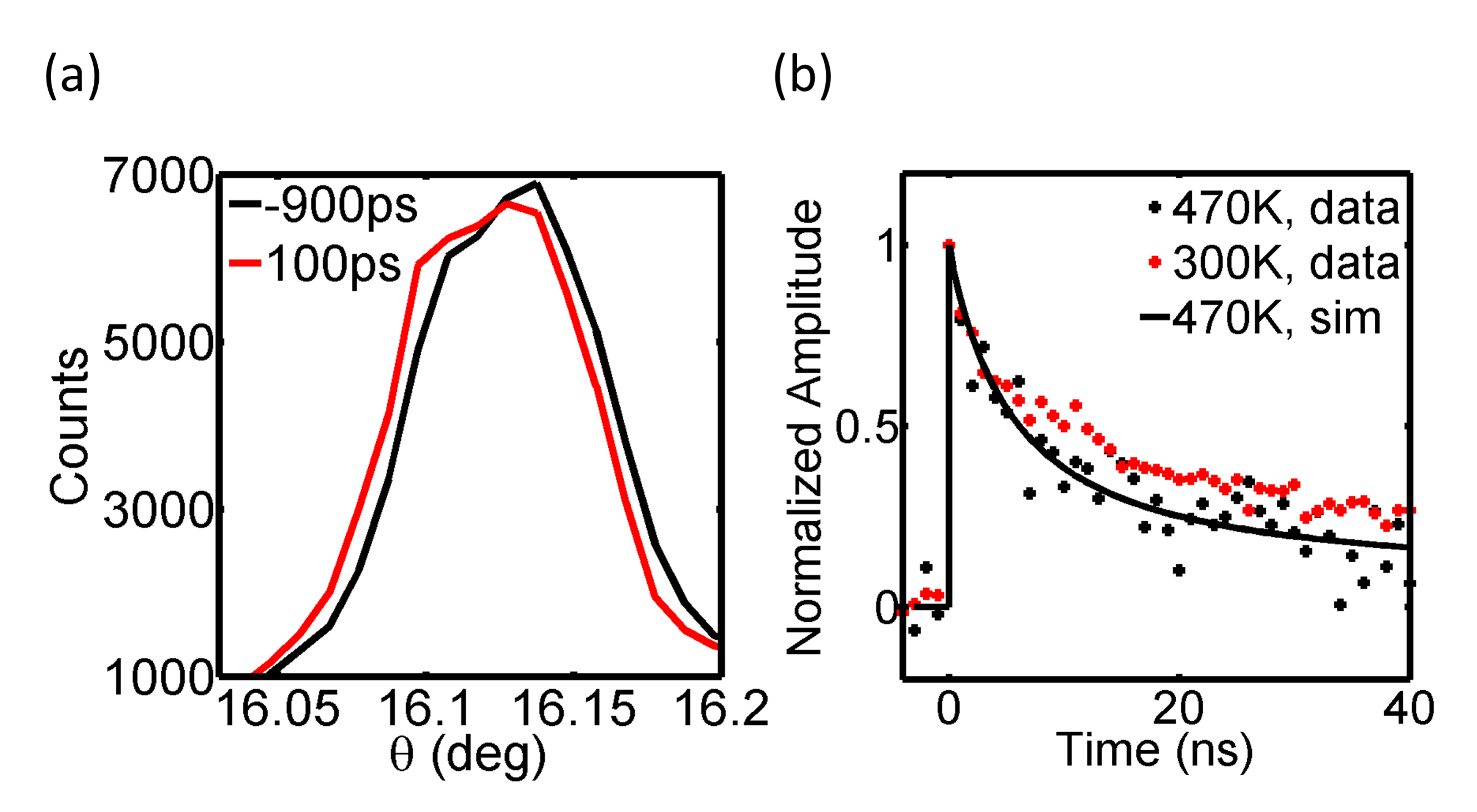}
\caption{Terahertz-pump x-ray probe measurements within split-ring resonator structure. (a) (002) rocking curve before (t=-900 ps) and after (t=100 ps) time zero showing large-amplitude field-induced tensile strain. (b) Nanosecond time-dependent intensity measured at the low angle side of the rocking curve below and above the Curie temperature. Also shown (solid line) are fits to a thermal model for the cooling of the BTO film in the paraelectric phase into the substrate, described in the Supplemental Materials.}
\label{Fig3}
\end{figure}

The observed changes in the diffracted intensity are consistent with rocking curve scans (measured by integrating over the entire diffracted intensity on an area detector) taken at different relative time delays between the THz and x-ray pulses. At t=5 ps, the differential rocking curve [Fig. \ref{Fig2}(e)] shows an overall increase at all angles probed.  At t = 15 ps, a clear shift of the rocking curve to lower angle emerges, with magnitude scaling quadratically with the applied field [Fig. 2(f)]. Measurements at higher fields within split-ring resonator structures using microfocused x-ray probes, shown in Figs. \ref{Fig3}(a,b), show larger effects, with peak shifts corresponding to terahertz-driven tensile strains of 0.04 $\%$ in the out-of-plane direction.  Based on static temperature-dependent x-ray diffraction measurements, we estimate that the observed shift towards lower angle corresponds to a field-induced heating of 72 K (see Supplemental Materials). The shift in rocking curve is a long-lived effect showing a 10 ns recovery [Fig. 3(b)]. This long recovery time constant is consistent with simulations of thermal transport assuming rough values for the interfacial thermal conductivity (see Supplemental Materials). Slight asymmetric changes of the rocking curve may be related to inhomogeneity in the strain within the film. Similar expansions in the c-axis lattice spacing are observed comparing THz-driven effects at room temperature to those observed above the Curie temperature [Fig. \ref{Fig3}(b)].   

We interpret the observed short- and long-time effects as a result of direct THz-driven coupling to the lattice, initially driving polar displacements along the direction of the applied electric field (within the plane of the sample).  Because measurements of the (003) diffraction peak provide sensitivity only to the out-of-plane atomic-displacements, the structure factor modulation observed at short times can therefore be understood as arising from the coupling of initially excited in-plane atomic displacements within the unit cell to out-of-plane displacements, as has been predicted to occur in prior theoretical work~\cite{Qi2009} and the MD simulation as discussed later. In particular, we show in the following that the observed increase in scattering intensity at short times is indicative of a transient increase in the out-of-plane RMS displacements of the central Ti atoms, associated with an ultrafast rotation of the macroscopic ferroelectric polarization. 
 
\begin{figure}
\includegraphics[width=\columnwidth]{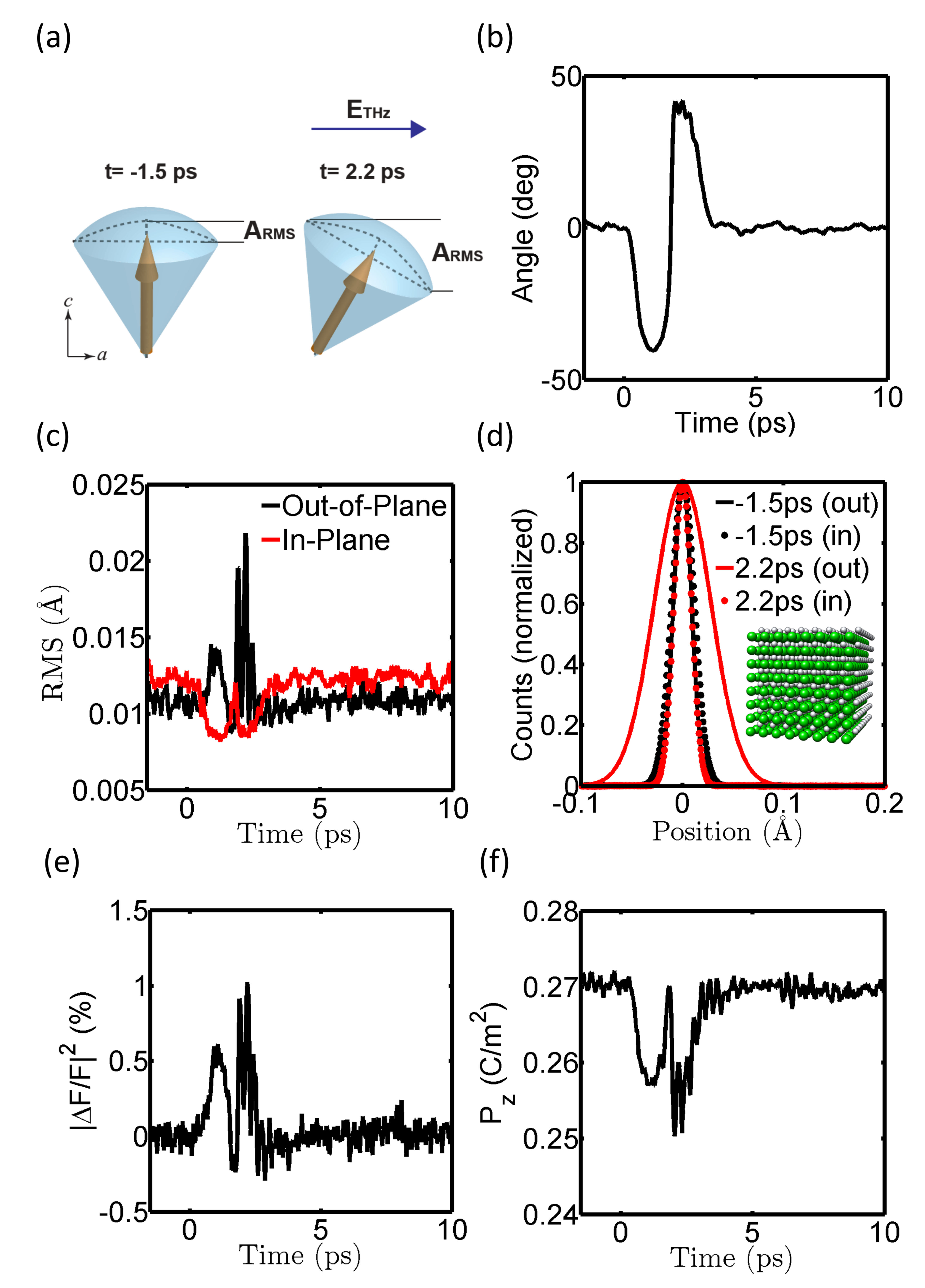}
\caption{MD simulation results. (a) Schematic showing how a rotation of the polarization leads to an increase in out-of-plane RMS displacements. b) The angle of the net polarization with respect to the c-axis. (c) The RMS displacement of the Ti atom along the in-plane and out-of-plane direction after in-plane THz excitation. (d) Corresponding histograms showing distributions of in-plane and out-of-plane displacements before and at the peak of the THz field. Inset shows MD simulaton supercell from which these histograms are calculated. (e) Calculated structure factor modulation for the (003) reflection.  (f) Projected net polarization along the c-axis as a function of time.  }
\label{Fig4}
\end{figure}

The structure factor for the probed (003) reflection can be written simply as
\begin{equation}
\sum_nf_ne^{2\pi i(hx_n+ky_n+lz_n)}|_{h=0,k=0,l=3} = f_{Ba}-f_{Ti}e^{-6\pi i\delta}
\end{equation}
where f$_{Ba}$ and f$_{Ti}$ are the scattering factors for the Ba and Ti atoms in the unit cell, (hkl) are the Miller indices for the probed reflection, $(x_n,y_n,z_n)$ are the fractional coordinates of the Ti atom, and $\delta$ is the out-of-plane fractional displacement of the  Ti atom measured from the center of the unit cell. Here we neglect the contribution from the oxygen atoms.  For small $\delta$, the scattered waves from the Ba and Ti atoms scatter $\pi$ out of phase with each other and destructively interfere, with the non-zero intensity of the reflection arising from the different scattering factors for the Ba and Ti atoms. Time-dependent displacements of the atoms will modulate this destructive interference and qualitatively explain the observed time-dependent increase in scattering efficiency.  In a simple harmonic oscillator model, if the THz pulse induces a time-dependent out-of-plane vibrational excitation of the Ti atom with amplitude A~\cite{Zamponi:2012dz}, the time-dependent structure factor becomes
\begin{equation}
F_{003}(t)=f_{Ba}-f_{Ti}\exp\left[-6\pi i(\delta+A(t))\right]
\end{equation}
Since no coherent oscillatory response is observed when spatially averaging over all unit cells, we consider the impact of a finite time-dependent RMS displacement $\sqrt{<A(t)^2>}=A_{RMS}(t)$. One may then show quantitatively from Equation 1 (see also Supplemental Materials, section 2) that this leads to a fractional increase in the (003) scattered intensity given by:
\begin{equation}
\frac{\Delta I}{I_o}\sim\frac{36f_{Ba}f_{Ti}\pi^2A_{RMS}^2(t)}{(f_{Ba}-f_{Ti})^2}
\end{equation}
For a 1$\%$ increase of the diffraction intensity as experimentally observed, Equation 3 gives a peak RMS displacement amplitude $A_{RMS}$=0.03 \r{A}, corresponding to an out-of-plane modulation comparable to the built-in ferroelectric displacement~\cite{Smith:2008kv}. We compare the observed response to molecular dynamics (MD) simulations (Fig. \ref{Fig4}, supplemental materials) which predict peak RMS displacements of the central Ti atom of 0.02 \r{A}.  From the calculated coordinates of each atom in the supercell (including the oxygen atoms)  we also calculate directly the time-dependent modulation in the structure factor, obtaining fractional modulations in the effective scattering intensity of 1$\%$ [Fig. \ref{Fig4}(e)] for the applied fields used here, in excellent agreement with the experimental results. We note this effect is in contrast to the ordinary decrease in scattering intensity that would be associated with a temperature-induced Debye-Waller effect.~\cite{Mannebach2015,Lindenberg2005}  Inclusion of the oxygen atoms in the above analytical calculation (See Supplemental Materials) gives a slightly revised equation which can be written as:
\begin{equation}
\frac{\Delta I}{I_o}\sim\frac{36\pi^2A_{RMS}^2(t)(f_{Ba}f_{Ti}+f_{Ba}f_{O}-4f_{Ti}f_O)}{(f_{Ba}-f_{Ti})^2+f_O^2-2f_{Ba}f_O+2f_{Ti}f_O}
\end{equation}
This changes the estimated magnitude of the induced RMS displacement by $\approx 10$\%.

Based on the observed THz-field-induced out-of-plane RMS displacement and the MD simulation, one can derive an atomistic understanding of structural response of a ferroelectric upon ultrafast electric field excitation, directly relating the induced out-of-plane motions to a field-induced transient polarization rotation of local dipoles following the applied in-plane THz field.  Before application of the THz field, BTO exhibits order-disorder character, with the Ti atoms distributed along the four $<111>$ directions of the unit cells within the tetragonal phase,~\cite{Chaves:1976wq, Zhang:2006wd, Comes:1968vx, Ravel:1998vl, Qi:2016} with this initial disorder likely underlying the lack of a coherent vibrational response. This is schematically shown in Fig. 4(a) (left) depicting the distribution of dipoles within the ensemble of unit cells probed. Upon THz excitation, these dipoles rotate in the transverse electric field of the bias pulse such that the intrinsic in-plane distribution is transformed to one with an increased component along the out-of-plane direction, shown as the enhanced RMS amplitude in Fig. 4(a) and in agreement with our observations. This simple model is supported by more detailed theoretical calculations: Fig. 4(b) shows calculations of the time-dependent induced rotation of the polarization in the applied THz field used in the experiments, showing rotation amplitudes of approximately 40 degrees. This magnitude is consistent with simple estimates based on the known transverse susceptibility of BTO.~\cite{Davis:2007gy}  Fig. 4(c) shows the time-dependence of the RMS displacements averaged over the simulated 10$\times$10$\times$10 supercell (shown in the inset to 4d, see also Supplemental Information) comparing the in-plane to out-of-plane response. This shows, consistent with the simple model described above, that the in-plane RMS distribution decreases whereas the out-of-plane increases.  The magnitude of the simulated increase in the out-of-plane displacements is in good agreement with our experimental measurement. In addition, Fig. 4(d) shows histograms of the simulated displacements comparing before excitation (t=-1.5 ps) and at the peak of the THz field (t=2.2 ps), showing a response similar to that shown in Fig. 4(c).  Finally, from the simulated spatial coordinates of the atoms within the supercell (Supplemental Materials, section 9), one may calculate directly the time-dependence of the (003) reflection, showing a transient increase and then decrease in Fig. 4(e), with good agreement comparing the magnitude of the change to the observed 
change in the diffracted intensity.  Both simulated and calculated structure factor modulations rise with a time-constant of a few picoseconds although the simulated response turns off on an about 3x faster time-scale, more closely following the applied field. From the perspective of the polarization rotation model described above, this recovery time-scale corresponds to the time for the polarization to reorient along its original direction and this effect may depend in a more complicated way on the thin-film geometry which is not accurately included in the simulations.  Further experiments are required to investigate this effect.  Although not directly measured here, the induced rotation leads to a decrease in the out-of-plane polarization of order 10$\%$ shown in Fig. 4(f), which can also be viewed from the perspective of a rotation of the ferroelectric polarization. This decrease corresponds to a displacement small compared to A$_{RMS}$ justifying the assumptions made above. We emphasize that the experimentally observed ultrafast recovery on picosecond time-scales shows that this time-dependent structure factor modulation and associated increase in RMS displacements are not associated with field-induced temperature jumps, which will recover on much slower (nanosecond) time-scales determined by thermal transport, as shown in Fig. 3.  Complementary measurements on the (012) reflection (with in-plane reciprocal lattice vector component perpendicular to the applied in-plane THz field), described in the Supplemental Materials, further confirm the above model. 

As noted above, the transient structure factor modulations are superimposed on a large amplitude and long-lived homogeneous shift of the diffraction peaks to lower angles, corresponding to an increase in the c-axis tetragonality. This response can be understood as arising from the initial vibrational excitation of the thin film, with the peak frequency of the THz pulse resonant with the $\sim 1$ THz Slater E(TO) mode in BTO~\cite{Hoshina2014}. We note that a detailed model including the epitaxial clamping of the film to the substrate indicates that the Bragg peak shift is associated with a  heating effect with magnitude $\approx$70 K at the highest applied fields. In particular this model accounts for the coupling between in-plane and out-of-plane stresses and the fact that the THz excitation only interacts to first order with the thin film and not the substrate. More details on this model are included in the Supplemental Materials Section 2.  This observed transient increase in tetragonality is consistent with a simple Joule heating model: Using known conductivities at 1 THz (dominated by vibrational degrees of freedom), the energy absorbed by the thin film leads to an estimated maximum temperature change $\Delta T\sim\frac{\sigma E^2\Delta t}{C}$ = 20 K for a peak field E=0.7 MV$/$cm within the film where the conductivity $\sigma$ is calculated from the known optical constants at 1 THz frequency~\cite{Misra2004,Hoshina2014}, C is the heat capacity, and $\Delta$t is the temporal width of the applied THz pulse.  We estimate the peak field within the film using finite difference time domain simulations (shown in Fig. \ref{Fig1} with more details in the Supplemental Materials). This is also consistent with the effect size scaling quadratically with the THz electric field, as experimentally observed [Fig. 2(f)]. We note that the observed linear scaling with the intensity of the THz pulse is inconsistent with THz-driven ionization processes~\cite{Hirori2011}. Further, this scaling and the observation of similar effects above the Curie temperature [Fig. 3(b)] rules out potential piezoelectric responses, while the long-lived nature of the response rules out both piezoelectric and electrostrictive stresses as playing important roles in the observed dynamics. A quantitative model for the THz-induced stress and strain together with a dynamical diffraction model~\cite{Larson1980, Wie1986, Schick:2014ie} (Supplemental Materials) captures fully the observed onset time and magnitude of the long-lived effects and allows for quantitative fitting of the observed response, also shown in Fig. 2.

To conclude, femtosecond x-ray diffraction measurements on THz-driven BTO directly capture the first atomic-scale steps in its electric-field-driven structural response. The combination of  experimental observations with first-principles-based MD simulations show evidence for large amplitude rotations of the ferroelectric polarization occurring on ps time-scales, as reflected in changes in the RMS out-of-plane displacements of the central Ti atom within the unit cell.  Additionally these measurements capture the coupling between in-plane and out-of-plane vibrational modes and the concomitant high-frequency acoustic strains that can be driven by electric field pulses.  The results are in good agreement with a model in which the THz field is directly coupled to low frequency modes of the BTO lattice and with first-principles-based MD simulations.  Future studies probing additional x-ray reflections may enable direct reconstruction of the full unit cell structural changes.  In this transverse geometry, additional opportunities exist for  visualization of dynamic electrocaloric responses and associated devices in which field-induced cooling or heating effects are directly resolved under both collinear (E $||$ P) and non-collinear (E$\perp$P) geometries. Novel possibilities with respect to terahertz-frequency photonic switches with unique photoelectromechanical responses also follow from this work.

This work was supported by the Department of Energy, Basic Energy Sciences, Materials Sciences and Engineering Division. A.R.D. acknowledges support from the Army Research Office under grant W911NF-14-1-0104. H.W. and L.W.M. acknowledge support from the Department of Energy under grant No. DE-SC0012375. Work at Argonne was supported by the U.S Department of Energy, Office of Science, Office of Basic Energy Sciences, under Contract No. DE-AC02-06CH11357.  S. L. acknowledges support from the US Department of Energy, Office of Basic Energy Sciences, under grant DE-FG02-07ER15920, as well as support from the Carnegie Institution for Science. Y.Q. acknowledges support from the National Science Foundation under grant CMMI-1334241. A.M.R. acknowledges support from the Office of Naval Research, under grant N00014-12-1-1033.  Research by the MIT group was supported in part by Office of Naval Research Grant No. N00014-13-1-0509 and National Science Foundation Grant No. CHE-1111557.  Use of the Linac Coherent Light Source (LCLS), SLAC National Accelerator Laboratory, is supported by the U.S. Department of Energy, Office of Science, Office of Basic Energy Sciences under Contract No. DE-AC02-76SF00515.


\begin{thebibliography}{41}%
\makeatletter
\providecommand \@ifxundefined [1]{%
 \@ifx{#1\undefined}
}%
\providecommand \@ifnum [1]{%
 \ifnum #1\expandafter \@firstoftwo
 \else \expandafter \@secondoftwo
 \fi
}%
\providecommand \@ifx [1]{%
 \ifx #1\expandafter \@firstoftwo
 \else \expandafter \@secondoftwo
 \fi
}%
\providecommand \natexlab [1]{#1}%
\providecommand \enquote  [1]{``#1''}%
\providecommand \bibnamefont  [1]{#1}%
\providecommand \bibfnamefont [1]{#1}%
\providecommand \citenamefont [1]{#1}%
\providecommand \href@noop [0]{\@secondoftwo}%
\providecommand \href [0]{\begingroup \@sanitize@url \@href}%
\providecommand \@href[1]{\@@startlink{#1}\@@href}%
\providecommand \@@href[1]{\endgroup#1\@@endlink}%
\providecommand \@sanitize@url [0]{\catcode `\\12\catcode `\$12\catcode
  `\&12\catcode `\#12\catcode `\^12\catcode `\_12\catcode `\%12\relax}%
\providecommand \@@startlink[1]{}%
\providecommand \@@endlink[0]{}%
\providecommand \url  [0]{\begingroup\@sanitize@url \@url }%
\providecommand \@url [1]{\endgroup\@href {#1}{\urlprefix }}%
\providecommand \urlprefix  [0]{URL }%
\providecommand \Eprint [0]{\href }%
\providecommand \doibase [0]{http://dx.doi.org/}%
\providecommand \selectlanguage [0]{\@gobble}%
\providecommand \bibinfo  [0]{\@secondoftwo}%
\providecommand \bibfield  [0]{\@secondoftwo}%
\providecommand \translation [1]{[#1]}%
\providecommand \BibitemOpen [0]{}%
\providecommand \bibitemStop [0]{}%
\providecommand \bibitemNoStop [0]{.\EOS\space}%
\providecommand \EOS [0]{\spacefactor3000\relax}%
\providecommand \BibitemShut  [1]{\csname bibitem#1\endcsname}%
\let\auto@bib@innerbib\@empty
\bibitem [{\citenamefont {Grinberg}\ \emph {et~al.}(2013)\citenamefont
  {Grinberg}, \citenamefont {West}, \citenamefont {Torres}, \citenamefont
  {Gou}, \citenamefont {Stein}, \citenamefont {Wu}, \citenamefont {Chen},
  \citenamefont {Gallo}, \citenamefont {Akbashev}, \citenamefont {Davies},
  \citenamefont {Spanier},\ and\ \citenamefont {Rappe}}]{Grinberg2013}%
  \BibitemOpen
  \bibfield  {author} {\bibinfo {author} {\bibfnamefont {I.}~\bibnamefont
  {Grinberg}}, \bibinfo {author} {\bibfnamefont {D.~V.}\ \bibnamefont {West}},
  \bibinfo {author} {\bibfnamefont {M.}~\bibnamefont {Torres}}, \bibinfo
  {author} {\bibfnamefont {G.}~\bibnamefont {Gou}}, \bibinfo {author}
  {\bibfnamefont {D.~M.}\ \bibnamefont {Stein}}, \bibinfo {author}
  {\bibfnamefont {L.}~\bibnamefont {Wu}}, \bibinfo {author} {\bibfnamefont
  {G.}~\bibnamefont {Chen}}, \bibinfo {author} {\bibfnamefont {E.~M.}\
  \bibnamefont {Gallo}}, \bibinfo {author} {\bibfnamefont {A.~R.}\ \bibnamefont
  {Akbashev}}, \bibinfo {author} {\bibfnamefont {P.~K.}\ \bibnamefont
  {Davies}}, \bibinfo {author} {\bibfnamefont {J.~E.}\ \bibnamefont {Spanier}},
  \ and\ \bibinfo {author} {\bibfnamefont {A.~M.}\ \bibnamefont {Rappe}},\
  }\href@noop {} {\bibfield  {journal} {\bibinfo  {journal} {Nature}\ }\textbf
  {\bibinfo {volume} {503}},\ \bibinfo {pages} {509} (\bibinfo {year}
  {2013})}\BibitemShut {NoStop}%
\bibitem [{\citenamefont {Choi}\ \emph {et~al.}(2009)\citenamefont {Choi},
  \citenamefont {Lee}, \citenamefont {Choi}, \citenamefont {Kiryukhin},\ and\
  \citenamefont {Cheong}}]{Choi2009}%
  \BibitemOpen
  \bibfield  {author} {\bibinfo {author} {\bibfnamefont {T.}~\bibnamefont
  {Choi}}, \bibinfo {author} {\bibfnamefont {S.}~\bibnamefont {Lee}}, \bibinfo
  {author} {\bibfnamefont {Y.~J.}\ \bibnamefont {Choi}}, \bibinfo {author}
  {\bibfnamefont {V.}~\bibnamefont {Kiryukhin}}, \ and\ \bibinfo {author}
  {\bibfnamefont {S.-W.}\ \bibnamefont {Cheong}},\ }\href@noop {} {\bibfield
  {journal} {\bibinfo  {journal} {Science}\ }\textbf {\bibinfo {volume}
  {324}},\ \bibinfo {pages} {63} (\bibinfo {year} {2009})}\BibitemShut
  {NoStop}%
\bibitem [{\citenamefont {Chen}\ \emph {et~al.}(2014)\citenamefont {Chen},
  \citenamefont {Chang}, \citenamefont {Zhang}, \citenamefont {Ok},
  \citenamefont {Ling}, \citenamefont {Mihnev}, \citenamefont {Norris},\ and\
  \citenamefont {Guo}}]{Chen2014}%
  \BibitemOpen
  \bibfield  {author} {\bibinfo {author} {\bibfnamefont {S.-L.}\ \bibnamefont
  {Chen}}, \bibinfo {author} {\bibfnamefont {Y.-C.}\ \bibnamefont {Chang}},
  \bibinfo {author} {\bibfnamefont {C.}~\bibnamefont {Zhang}}, \bibinfo
  {author} {\bibfnamefont {J.~G.}\ \bibnamefont {Ok}}, \bibinfo {author}
  {\bibfnamefont {T.}~\bibnamefont {Ling}}, \bibinfo {author} {\bibfnamefont
  {M.~T.}\ \bibnamefont {Mihnev}}, \bibinfo {author} {\bibfnamefont {T.~B.}\
  \bibnamefont {Norris}}, \ and\ \bibinfo {author} {\bibfnamefont {L.~J.}\
  \bibnamefont {Guo}},\ }\href@noop {} {\bibfield  {journal} {\bibinfo
  {journal} {Nat. Photon.}\ }\textbf {\bibinfo {volume} {8}},\ \bibinfo {pages}
  {537} (\bibinfo {year} {2014})}\BibitemShut {NoStop}%
\bibitem [{\citenamefont {Yang}\ \emph {et~al.}(2010)\citenamefont {Yang},
  \citenamefont {Seidel}, \citenamefont {Byrnes} \emph {et~al.}}]{Yang2010}%
  \BibitemOpen
  \bibfield  {author} {\bibinfo {author} {\bibfnamefont {S.~Y.}\ \bibnamefont
  {Yang}}, \bibinfo {author} {\bibfnamefont {J.}~\bibnamefont {Seidel}},
  \bibinfo {author} {\bibfnamefont {S.~J.}\ \bibnamefont {Byrnes}},  \emph
  {et~al.},\ }\href@noop {} {\bibfield  {journal} {\bibinfo  {journal} {Nat.
  Nanotech.}\ }\textbf {\bibinfo {volume} {5}},\ \bibinfo {pages} {143}
  (\bibinfo {year} {2010})}\BibitemShut {NoStop}%
\bibitem [{\citenamefont {Young}\ and\ \citenamefont
  {Rappe}(2012)}]{Young2012}%
  \BibitemOpen
  \bibfield  {author} {\bibinfo {author} {\bibfnamefont {S.~M.}\ \bibnamefont
  {Young}}\ and\ \bibinfo {author} {\bibfnamefont {A.~M.}\ \bibnamefont
  {Rappe}},\ }\href@noop {} {\bibfield  {journal} {\bibinfo  {journal} {Phys.
  Rev. Lett.}\ }\textbf {\bibinfo {volume} {109}},\ \bibinfo {pages} {116601}
  (\bibinfo {year} {2012})}\BibitemShut {NoStop}%
\bibitem [{\citenamefont {Schmising}\ \emph {et~al.}(2007)\citenamefont
  {Schmising}, \citenamefont {Bargheer}, \citenamefont {Kiel},\ and\
  \citenamefont {Zhavoronkov}}]{KorffSchmising2007}%
  \BibitemOpen
  \bibfield  {author} {\bibinfo {author} {\bibfnamefont {C.~K.}\ \bibnamefont
  {Schmising}}, \bibinfo {author} {\bibfnamefont {M.}~\bibnamefont {Bargheer}},
  \bibinfo {author} {\bibfnamefont {M.}~\bibnamefont {Kiel}}, \ and\ \bibinfo
  {author} {\bibfnamefont {N.}~\bibnamefont {Zhavoronkov}},\ }\href@noop {}
  {\bibfield  {journal} {\bibinfo  {journal} {Phys. Rev. Lett.}\ }\textbf
  {\bibinfo {volume} {98}},\ \bibinfo {pages} {257601} (\bibinfo {year}
  {2007})}\BibitemShut {NoStop}%
\bibitem [{\citenamefont {Miyamoto}\ \emph {et~al.}(2013)\citenamefont
  {Miyamoto}, \citenamefont {Yada}, \citenamefont {Yamakawa},\ and\
  \citenamefont {Okamoto}}]{Miyamoto2013}%
  \BibitemOpen
  \bibfield  {author} {\bibinfo {author} {\bibfnamefont {T.}~\bibnamefont
  {Miyamoto}}, \bibinfo {author} {\bibfnamefont {H.}~\bibnamefont {Yada}},
  \bibinfo {author} {\bibfnamefont {H.}~\bibnamefont {Yamakawa}}, \ and\
  \bibinfo {author} {\bibfnamefont {H.}~\bibnamefont {Okamoto}},\ }\href@noop
  {} {\bibfield  {journal} {\bibinfo  {journal} {Nat. Commun.}\ }\textbf
  {\bibinfo {volume} {4}},\ \bibinfo {pages} {2586} (\bibinfo {year}
  {2013})}\BibitemShut {NoStop}%
\bibitem [{\citenamefont {Wen}\ \emph {et~al.}(2013)\citenamefont {Wen},
  \citenamefont {Chen}, \citenamefont {Cosgriff} \emph {et~al.}}]{Wen2013}%
  \BibitemOpen
  \bibfield  {author} {\bibinfo {author} {\bibfnamefont {H.}~\bibnamefont
  {Wen}}, \bibinfo {author} {\bibfnamefont {P.}~\bibnamefont {Chen}}, \bibinfo
  {author} {\bibfnamefont {M.~P.}\ \bibnamefont {Cosgriff}},  \emph {et~al.},\
  }\href@noop {} {\bibfield  {journal} {\bibinfo  {journal} {Phys. Rev. Lett.}\
  }\textbf {\bibinfo {volume} {110}},\ \bibinfo {pages} {037601} (\bibinfo
  {year} {2013})}\BibitemShut {NoStop}%
\bibitem [{\citenamefont {Daranciang}\ \emph {et~al.}(2012)\citenamefont
  {Daranciang}, \citenamefont {Highland}, \citenamefont {Wen} \emph
  {et~al.}}]{Daranciang2012}%
  \BibitemOpen
  \bibfield  {author} {\bibinfo {author} {\bibfnamefont {D.}~\bibnamefont
  {Daranciang}}, \bibinfo {author} {\bibfnamefont {M.~J.}\ \bibnamefont
  {Highland}}, \bibinfo {author} {\bibfnamefont {H.}~\bibnamefont {Wen}},
  \emph {et~al.},\ }\href@noop {} {\bibfield  {journal} {\bibinfo  {journal}
  {Phys. Rev. Lett.}\ }\textbf {\bibinfo {volume} {108}},\ \bibinfo {pages}
  {087601} (\bibinfo {year} {2012})}\BibitemShut {NoStop}%
\bibitem [{\citenamefont {Kampfrath}\ \emph {et~al.}(2013)\citenamefont
  {Kampfrath}, \citenamefont {Tanaka},\ and\ \citenamefont
  {Nelson}}]{Kampfrath2013}%
  \BibitemOpen
  \bibfield  {author} {\bibinfo {author} {\bibfnamefont {T.}~\bibnamefont
  {Kampfrath}}, \bibinfo {author} {\bibfnamefont {K.}~\bibnamefont {Tanaka}}, \
  and\ \bibinfo {author} {\bibfnamefont {K.~A.}\ \bibnamefont {Nelson}},\
  }\href@noop {} {\bibfield  {journal} {\bibinfo  {journal} {Nat. Photon.}\
  }\textbf {\bibinfo {volume} {7}},\ \bibinfo {pages} {680} (\bibinfo {year}
  {2013})}\BibitemShut {NoStop}%
\bibitem [{\citenamefont {Chen}\ \emph {et~al.}(2015)\citenamefont {Chen},
  \citenamefont {Goodfellow}, \citenamefont {Liu}, \citenamefont {Grinberg},
  \citenamefont {Hoffmann}, \citenamefont {Damodaran}, \citenamefont {Zhu},
  \citenamefont {Zalden}, \citenamefont {Zhang}, \citenamefont {Takeuchi},
  \citenamefont {Rappe}, \citenamefont {Martin}, \citenamefont {Wen},\ and\
  \citenamefont {Lindenberg}}]{Chen2015}%
  \BibitemOpen
  \bibfield  {author} {\bibinfo {author} {\bibfnamefont {F.}~\bibnamefont
  {Chen}}, \bibinfo {author} {\bibfnamefont {J.}~\bibnamefont {Goodfellow}},
  \bibinfo {author} {\bibfnamefont {S.}~\bibnamefont {Liu}}, \bibinfo {author}
  {\bibfnamefont {I.}~\bibnamefont {Grinberg}}, \bibinfo {author}
  {\bibfnamefont {M.~C.}\ \bibnamefont {Hoffmann}}, \bibinfo {author}
  {\bibfnamefont {A.~R.}\ \bibnamefont {Damodaran}}, \bibinfo {author}
  {\bibfnamefont {Y.}~\bibnamefont {Zhu}}, \bibinfo {author} {\bibfnamefont
  {P.}~\bibnamefont {Zalden}}, \bibinfo {author} {\bibfnamefont
  {X.}~\bibnamefont {Zhang}}, \bibinfo {author} {\bibfnamefont
  {I.}~\bibnamefont {Takeuchi}}, \bibinfo {author} {\bibfnamefont {A.~M.}\
  \bibnamefont {Rappe}}, \bibinfo {author} {\bibfnamefont {L.~W.}\ \bibnamefont
  {Martin}}, \bibinfo {author} {\bibfnamefont {H.}~\bibnamefont {Wen}}, \ and\
  \bibinfo {author} {\bibfnamefont {A.~M.}\ \bibnamefont {Lindenberg}},\
  }\href@noop {} {\bibfield  {journal} {\bibinfo  {journal} {Adv. Mater.}\
  }\textbf {\bibinfo {volume} {27}},\ \bibinfo {pages} {6371} (\bibinfo {year}
  {2015})}\BibitemShut {NoStop}%
\bibitem [{\citenamefont {Kubacka}\ \emph {et~al.}(2014)\citenamefont
  {Kubacka}, \citenamefont {Johnson}, \citenamefont {Hoffmann} \emph
  {et~al.}}]{Kubacka2014}%
  \BibitemOpen
  \bibfield  {author} {\bibinfo {author} {\bibfnamefont {T.}~\bibnamefont
  {Kubacka}}, \bibinfo {author} {\bibfnamefont {J.~A.}\ \bibnamefont
  {Johnson}}, \bibinfo {author} {\bibfnamefont {M.~C.}\ \bibnamefont
  {Hoffmann}},  \emph {et~al.},\ }\href@noop {} {\bibfield  {journal} {\bibinfo
   {journal} {Science}\ }\textbf {\bibinfo {volume} {343}},\ \bibinfo {pages}
  {1333} (\bibinfo {year} {2014})}\BibitemShut {NoStop}%
\bibitem [{\citenamefont {Kundys}\ \emph {et~al.}(2010)\citenamefont {Kundys},
  \citenamefont {Viret}, \citenamefont {Colson},\ and\ \citenamefont
  {Kundys}}]{Kundys2010}%
  \BibitemOpen
  \bibfield  {author} {\bibinfo {author} {\bibfnamefont {B.}~\bibnamefont
  {Kundys}}, \bibinfo {author} {\bibfnamefont {M.}~\bibnamefont {Viret}},
  \bibinfo {author} {\bibfnamefont {D.}~\bibnamefont {Colson}}, \ and\ \bibinfo
  {author} {\bibfnamefont {D.}~\bibnamefont {Kundys}},\ }\href@noop {}
  {\bibfield  {journal} {\bibinfo  {journal} {Nat. Mater.}\ }\textbf {\bibinfo
  {volume} {9}},\ \bibinfo {pages} {803} (\bibinfo {year} {2010})}\BibitemShut
  {NoStop}%
\bibitem [{\citenamefont {Shin}\ \emph {et~al.}(2007)\citenamefont {Shin},
  \citenamefont {Grinberg}, \citenamefont {Chen},\ and\ \citenamefont
  {Rappe}}]{Shin2007}%
  \BibitemOpen
  \bibfield  {author} {\bibinfo {author} {\bibfnamefont {Y.-H.}\ \bibnamefont
  {Shin}}, \bibinfo {author} {\bibfnamefont {I.}~\bibnamefont {Grinberg}},
  \bibinfo {author} {\bibfnamefont {I.-W.}\ \bibnamefont {Chen}}, \ and\
  \bibinfo {author} {\bibfnamefont {A.~M.}\ \bibnamefont {Rappe}},\ }\href@noop
  {} {\bibfield  {journal} {\bibinfo  {journal} {Nature}\ }\textbf {\bibinfo
  {volume} {449}},\ \bibinfo {pages} {881} (\bibinfo {year}
  {2007})}\BibitemShut {NoStop}%
\bibitem [{\citenamefont {Grigoriev}\ \emph {et~al.}(2006)\citenamefont
  {Grigoriev}, \citenamefont {Do}, \citenamefont {Kim}, \citenamefont {Eom},
  \citenamefont {Adams}, \citenamefont {Dufresne},\ and\ \citenamefont
  {Evans}}]{Grigoriev2006}%
  \BibitemOpen
  \bibfield  {author} {\bibinfo {author} {\bibfnamefont {A.}~\bibnamefont
  {Grigoriev}}, \bibinfo {author} {\bibfnamefont {D.-H.}\ \bibnamefont {Do}},
  \bibinfo {author} {\bibfnamefont {D.~M.}\ \bibnamefont {Kim}}, \bibinfo
  {author} {\bibfnamefont {C.-B.}\ \bibnamefont {Eom}}, \bibinfo {author}
  {\bibfnamefont {B.}~\bibnamefont {Adams}}, \bibinfo {author} {\bibfnamefont
  {E.~M.}\ \bibnamefont {Dufresne}}, \ and\ \bibinfo {author} {\bibfnamefont
  {P.~G.}\ \bibnamefont {Evans}},\ }\href@noop {} {\bibfield  {journal}
  {\bibinfo  {journal} {Phys. Rev. Lett.}\ }\textbf {\bibinfo {volume} {96}},\
  \bibinfo {pages} {187601} (\bibinfo {year} {2006})}\BibitemShut {NoStop}%
\bibitem [{\citenamefont {Jiang}\ \emph {et~al.}(2012)\citenamefont {Jiang},
  \citenamefont {Lee}, \citenamefont {Hwang},\ and\ \citenamefont
  {Scott}}]{Jiang2012}%
  \BibitemOpen
  \bibfield  {author} {\bibinfo {author} {\bibfnamefont {A.~Q.}\ \bibnamefont
  {Jiang}}, \bibinfo {author} {\bibfnamefont {H.~J.}\ \bibnamefont {Lee}},
  \bibinfo {author} {\bibfnamefont {C.~S.}\ \bibnamefont {Hwang}}, \ and\
  \bibinfo {author} {\bibfnamefont {J.~F.}\ \bibnamefont {Scott}},\ }\href@noop
  {} {\bibfield  {journal} {\bibinfo  {journal} {Adv. Funct. Mater.}\ }\textbf
  {\bibinfo {volume} {22}},\ \bibinfo {pages} {192} (\bibinfo {year}
  {2012})}\BibitemShut {NoStop}%
\bibitem [{\citenamefont {Grinberg}\ \emph {et~al.}(2009)\citenamefont
  {Grinberg}, \citenamefont {Shin},\ and\ \citenamefont
  {Rappe}}]{Grinberg2009}%
  \BibitemOpen
  \bibfield  {author} {\bibinfo {author} {\bibfnamefont {I.}~\bibnamefont
  {Grinberg}}, \bibinfo {author} {\bibfnamefont {Y.~H.}\ \bibnamefont {Shin}},
  \ and\ \bibinfo {author} {\bibfnamefont {A.~M.}\ \bibnamefont {Rappe}},\
  }\href@noop {} {\bibfield  {journal} {\bibinfo  {journal} {Phys. Rev. Lett.}\
  }\textbf {\bibinfo {volume} {103}},\ \bibinfo {pages} {197601} (\bibinfo
  {year} {2009})}\BibitemShut {NoStop}%
\bibitem [{\citenamefont {Zenkevich}\ \emph {et~al.}(2014)\citenamefont
  {Zenkevich}, \citenamefont {Matveyev}, \citenamefont {Maksimova},
  \citenamefont {Gaynutdinov}, \citenamefont {Tolstikhina},\ and\ \citenamefont
  {Fridkin}}]{Zenkevich2014}%
  \BibitemOpen
  \bibfield  {author} {\bibinfo {author} {\bibfnamefont {A.}~\bibnamefont
  {Zenkevich}}, \bibinfo {author} {\bibfnamefont {Y.}~\bibnamefont {Matveyev}},
  \bibinfo {author} {\bibfnamefont {K.}~\bibnamefont {Maksimova}}, \bibinfo
  {author} {\bibfnamefont {R.}~\bibnamefont {Gaynutdinov}}, \bibinfo {author}
  {\bibfnamefont {A.}~\bibnamefont {Tolstikhina}}, \ and\ \bibinfo {author}
  {\bibfnamefont {V.}~\bibnamefont {Fridkin}},\ }\href@noop {} {\bibfield
  {journal} {\bibinfo  {journal} {Phys. Rev. B}\ }\textbf {\bibinfo {volume}
  {90}},\ \bibinfo {pages} {161409} (\bibinfo {year} {2014})}\BibitemShut
  {NoStop}%
\bibitem [{\citenamefont {Li}\ \emph {et~al.}(2004)\citenamefont {Li},
  \citenamefont {Nagaraj}, \citenamefont {Liang}, \citenamefont {Cao},
  \citenamefont {Lee},\ and\ \citenamefont {Ramesh}}]{Li2004}%
  \BibitemOpen
  \bibfield  {author} {\bibinfo {author} {\bibfnamefont {J.}~\bibnamefont
  {Li}}, \bibinfo {author} {\bibfnamefont {B.}~\bibnamefont {Nagaraj}},
  \bibinfo {author} {\bibfnamefont {H.}~\bibnamefont {Liang}}, \bibinfo
  {author} {\bibfnamefont {W.}~\bibnamefont {Cao}}, \bibinfo {author}
  {\bibfnamefont {C.~H.}\ \bibnamefont {Lee}}, \ and\ \bibinfo {author}
  {\bibfnamefont {R.}~\bibnamefont {Ramesh}},\ }\href@noop {} {\bibfield
  {journal} {\bibinfo  {journal} {Appl. Phys. Lett.}\ }\textbf {\bibinfo
  {volume} {84}},\ \bibinfo {pages} {1174} (\bibinfo {year}
  {2004})}\BibitemShut {NoStop}%
\bibitem [{\citenamefont {Dougherty}\ \emph {et~al.}(1994)\citenamefont
  {Dougherty}, \citenamefont {Wiederrecht}, \citenamefont {Nelson},
  \citenamefont {Garrett}, \citenamefont {Jenssen},\ and\ \citenamefont
  {Warde}}]{Dougherty1994}%
  \BibitemOpen
  \bibfield  {author} {\bibinfo {author} {\bibfnamefont {T.~P.}\ \bibnamefont
  {Dougherty}}, \bibinfo {author} {\bibfnamefont {G.~P.}\ \bibnamefont
  {Wiederrecht}}, \bibinfo {author} {\bibfnamefont {K.~A.}\ \bibnamefont
  {Nelson}}, \bibinfo {author} {\bibfnamefont {M.~H.}\ \bibnamefont {Garrett}},
  \bibinfo {author} {\bibfnamefont {H.~P.}\ \bibnamefont {Jenssen}}, \ and\
  \bibinfo {author} {\bibfnamefont {C.}~\bibnamefont {Warde}},\ }\href@noop {}
  {\bibfield  {journal} {\bibinfo  {journal} {Phys. Rev. B}\ }\textbf {\bibinfo
  {volume} {50}},\ \bibinfo {pages} {8996} (\bibinfo {year}
  {1994})}\BibitemShut {NoStop}%
\bibitem [{\citenamefont {Katayama}\ \emph {et~al.}(2012)\citenamefont
  {Katayama}, \citenamefont {Aoki}, \citenamefont {Takeda}, \citenamefont
  {Shimosato}, \citenamefont {Ashida}, \citenamefont {Kinjo}, \citenamefont
  {Kawayama}, \citenamefont {Tonouchi}, \citenamefont {Nagai},\ and\
  \citenamefont {Tanaka}}]{Katayama2012}%
  \BibitemOpen
  \bibfield  {author} {\bibinfo {author} {\bibfnamefont {I.}~\bibnamefont
  {Katayama}}, \bibinfo {author} {\bibfnamefont {H.}~\bibnamefont {Aoki}},
  \bibinfo {author} {\bibfnamefont {J.}~\bibnamefont {Takeda}}, \bibinfo
  {author} {\bibfnamefont {H.}~\bibnamefont {Shimosato}}, \bibinfo {author}
  {\bibfnamefont {M.}~\bibnamefont {Ashida}}, \bibinfo {author} {\bibfnamefont
  {R.}~\bibnamefont {Kinjo}}, \bibinfo {author} {\bibfnamefont
  {I.}~\bibnamefont {Kawayama}}, \bibinfo {author} {\bibfnamefont
  {M.}~\bibnamefont {Tonouchi}}, \bibinfo {author} {\bibfnamefont
  {M.}~\bibnamefont {Nagai}}, \ and\ \bibinfo {author} {\bibfnamefont
  {K.}~\bibnamefont {Tanaka}},\ }\href@noop {} {\bibfield  {journal} {\bibinfo
  {journal} {Phys. Rev. Lett.}\ }\textbf {\bibinfo {volume} {108}},\ \bibinfo
  {pages} {097401} (\bibinfo {year} {2012})}\BibitemShut {NoStop}%
\bibitem [{\citenamefont {Istomin}\ \emph {et~al.}(2007)\citenamefont
  {Istomin}, \citenamefont {Kotaidis}, \citenamefont {Plech},\ and\
  \citenamefont {Kong}}]{Istomin2007}%
  \BibitemOpen
  \bibfield  {author} {\bibinfo {author} {\bibfnamefont {K.}~\bibnamefont
  {Istomin}}, \bibinfo {author} {\bibfnamefont {V.}~\bibnamefont {Kotaidis}},
  \bibinfo {author} {\bibfnamefont {A.}~\bibnamefont {Plech}}, \ and\ \bibinfo
  {author} {\bibfnamefont {Q.}~\bibnamefont {Kong}},\ }\href@noop {} {\bibfield
   {journal} {\bibinfo  {journal} {Appl. Phys. Lett.}\ }\textbf {\bibinfo
  {volume} {90}},\ \bibinfo {pages} {022905} (\bibinfo {year}
  {2007})}\BibitemShut {NoStop}%
\bibitem [{\citenamefont {Zamponi}\ \emph {et~al.}(2012)\citenamefont
  {Zamponi}, \citenamefont {Rothhardt}, \citenamefont {Stingl}, \citenamefont
  {Woerner},\ and\ \citenamefont {Elsaesser}}]{Zamponi:2012dz}%
  \BibitemOpen
  \bibfield  {author} {\bibinfo {author} {\bibfnamefont {F.}~\bibnamefont
  {Zamponi}}, \bibinfo {author} {\bibfnamefont {P.}~\bibnamefont {Rothhardt}},
  \bibinfo {author} {\bibfnamefont {J.}~\bibnamefont {Stingl}}, \bibinfo
  {author} {\bibfnamefont {M.}~\bibnamefont {Woerner}}, \ and\ \bibinfo
  {author} {\bibfnamefont {T.}~\bibnamefont {Elsaesser}},\ }\href@noop {}
  {\bibfield  {journal} {\bibinfo  {journal} {Proc. Natl. Acad. Sci.}\ }\textbf
  {\bibinfo {volume} {109}},\ \bibinfo {pages} {5207} (\bibinfo {year}
  {2012})}\BibitemShut {NoStop}%
\bibitem [{\citenamefont {Qi}\ \emph {et~al.}(2009)\citenamefont {Qi},
  \citenamefont {Shin}, \citenamefont {Yeh}, \citenamefont {Nelson},\ and\
  \citenamefont {Rappe}}]{Qi2009}%
  \BibitemOpen
  \bibfield  {author} {\bibinfo {author} {\bibfnamefont {T.}~\bibnamefont
  {Qi}}, \bibinfo {author} {\bibfnamefont {Y.-H.}\ \bibnamefont {Shin}},
  \bibinfo {author} {\bibfnamefont {K.-L.}\ \bibnamefont {Yeh}}, \bibinfo
  {author} {\bibfnamefont {K.~A.}\ \bibnamefont {Nelson}}, \ and\ \bibinfo
  {author} {\bibfnamefont {A.~M.}\ \bibnamefont {Rappe}},\ }\href@noop {}
  {\bibfield  {journal} {\bibinfo  {journal} {Phys. Rev. Lett.}\ }\textbf
  {\bibinfo {volume} {102}},\ \bibinfo {pages} {247603} (\bibinfo {year}
  {2009})}\BibitemShut {NoStop}%
\bibitem [{\citenamefont {Zhu}\ \emph {et~al.}(2016)\citenamefont {Zhu},
  \citenamefont {Cai}, \citenamefont {Chen} \emph {et~al.}}]{Zhu:2016cp}%
  \BibitemOpen
  \bibfield  {author} {\bibinfo {author} {\bibfnamefont {Y.}~\bibnamefont
  {Zhu}}, \bibinfo {author} {\bibfnamefont {Z.}~\bibnamefont {Cai}}, \bibinfo
  {author} {\bibfnamefont {P.}~\bibnamefont {Chen}},  \emph {et~al.},\
  }\href@noop {} {\bibfield  {journal} {\bibinfo  {journal} {Sci. Rep.}\
  }\textbf {\bibinfo {volume} {6}},\ \bibinfo {pages} {21999} (\bibinfo {year}
  {2016})}\BibitemShut {NoStop}%
\bibitem [{\citenamefont {Nicoul}\ \emph {et~al.}(2011)\citenamefont {Nicoul},
  \citenamefont {Shymanovich}, \citenamefont {Tarasevitch}, \citenamefont
  {von~der Linde},\ and\ \citenamefont {Sokolowski-Tinten}}]{Nicoul2011}%
  \BibitemOpen
  \bibfield  {author} {\bibinfo {author} {\bibfnamefont {M.}~\bibnamefont
  {Nicoul}}, \bibinfo {author} {\bibfnamefont {U.}~\bibnamefont {Shymanovich}},
  \bibinfo {author} {\bibfnamefont {A.}~\bibnamefont {Tarasevitch}}, \bibinfo
  {author} {\bibfnamefont {D.}~\bibnamefont {von~der Linde}}, \ and\ \bibinfo
  {author} {\bibfnamefont {K.}~\bibnamefont {Sokolowski-Tinten}},\ }\href@noop
  {} {\bibfield  {journal} {\bibinfo  {journal} {Appl. Phys. Lett.}\ }\textbf
  {\bibinfo {volume} {98}},\ \bibinfo {pages} {191902} (\bibinfo {year}
  {2011})}\BibitemShut {NoStop}%
\bibitem [{\citenamefont {Smith}\ \emph {et~al.}(2008)\citenamefont {Smith},
  \citenamefont {Page}, \citenamefont {Siegrist}, \citenamefont {Redmond},
  \citenamefont {Walter}, \citenamefont {Seshadri}, \citenamefont {Brus},\ and\
  \citenamefont {Steigerwald}}]{Smith:2008kv}%
  \BibitemOpen
  \bibfield  {author} {\bibinfo {author} {\bibfnamefont {M.~B.}\ \bibnamefont
  {Smith}}, \bibinfo {author} {\bibfnamefont {K.}~\bibnamefont {Page}},
  \bibinfo {author} {\bibfnamefont {T.}~\bibnamefont {Siegrist}}, \bibinfo
  {author} {\bibfnamefont {P.~L.}\ \bibnamefont {Redmond}}, \bibinfo {author}
  {\bibfnamefont {E.~C.}\ \bibnamefont {Walter}}, \bibinfo {author}
  {\bibfnamefont {R.}~\bibnamefont {Seshadri}}, \bibinfo {author}
  {\bibfnamefont {L.~E.}\ \bibnamefont {Brus}}, \ and\ \bibinfo {author}
  {\bibfnamefont {M.~L.}\ \bibnamefont {Steigerwald}},\ }\href@noop {}
  {\bibfield  {journal} {\bibinfo  {journal} {J. Am. Chem. Soc.}\ }\textbf
  {\bibinfo {volume} {130}},\ \bibinfo {pages} {6955} (\bibinfo {year}
  {2008})}\BibitemShut {NoStop}%
\bibitem [{\citenamefont {Mannebach}\ \emph {et~al.}(2015)\citenamefont
  {Mannebach}, \citenamefont {Li}, \citenamefont {Duerloo} \emph
  {et~al.}}]{Mannebach2015}%
  \BibitemOpen
  \bibfield  {author} {\bibinfo {author} {\bibfnamefont {E.~M.}\ \bibnamefont
  {Mannebach}}, \bibinfo {author} {\bibfnamefont {R.}~\bibnamefont {Li}},
  \bibinfo {author} {\bibfnamefont {K.-A.}\ \bibnamefont {Duerloo}},  \emph
  {et~al.},\ }\href@noop {} {\bibfield  {journal} {\bibinfo  {journal} {Nano
  Lett.}\ }\textbf {\bibinfo {volume} {15}},\ \bibinfo {pages} {6889} (\bibinfo
  {year} {2015})}\BibitemShut {NoStop}%
\bibitem [{\citenamefont {Lindenberg}\ \emph {et~al.}(2005)\citenamefont
  {Lindenberg} \emph {et~al.}}]{Lindenberg2005}%
  \BibitemOpen
  \bibfield  {author} {\bibinfo {author} {\bibfnamefont {A.~M.}\ \bibnamefont
  {Lindenberg}} \emph {et~al.},\ }\href@noop {} {\bibfield  {journal} {\bibinfo
   {journal} {Science}\ }\textbf {\bibinfo {volume} {308}},\ \bibinfo {pages}
  {392} (\bibinfo {year} {2005})}\BibitemShut {NoStop}%
\bibitem [{\citenamefont {Chaves}\ \emph {et~al.}(1976)\citenamefont {Chaves},
  \citenamefont {Barreto}, \citenamefont {Nogueira},\ and\ \citenamefont
  {Zeks}}]{Chaves:1976wq}%
  \BibitemOpen
  \bibfield  {author} {\bibinfo {author} {\bibfnamefont {A.~S.}\ \bibnamefont
  {Chaves}}, \bibinfo {author} {\bibfnamefont {F.}~\bibnamefont {Barreto}},
  \bibinfo {author} {\bibfnamefont {R.~A.}\ \bibnamefont {Nogueira}}, \ and\
  \bibinfo {author} {\bibfnamefont {B.}~\bibnamefont {Zeks}},\ }\href@noop {}
  {\bibfield  {journal} {\bibinfo  {journal} {Phys. Rev. B}\ }\textbf {\bibinfo
  {volume} {13}},\ \bibinfo {pages} {207} (\bibinfo {year} {1976})}\BibitemShut
  {NoStop}%
\bibitem [{\citenamefont {Zhang}\ \emph {et~al.}(2006)\citenamefont {Zhang},
  \citenamefont {Cagin},\ and\ \citenamefont {Goddard}}]{Zhang:2006wd}%
  \BibitemOpen
  \bibfield  {author} {\bibinfo {author} {\bibfnamefont {Q.}~\bibnamefont
  {Zhang}}, \bibinfo {author} {\bibfnamefont {T.}~\bibnamefont {Cagin}}, \ and\
  \bibinfo {author} {\bibfnamefont {W.~A.}\ \bibnamefont {Goddard}},\
  }\href@noop {} {\bibfield  {journal} {\bibinfo  {journal} {Proc. Natl. Acad.
  Sci.}\ }\textbf {\bibinfo {volume} {103}},\ \bibinfo {pages} {14695}
  (\bibinfo {year} {2006})}\BibitemShut {NoStop}%
\bibitem [{\citenamefont {Comes}\ \emph {et~al.}(1968)\citenamefont {Comes},
  \citenamefont {Lambert},\ and\ \citenamefont {Guinier}}]{Comes:1968vx}%
  \BibitemOpen
  \bibfield  {author} {\bibinfo {author} {\bibfnamefont {R.}~\bibnamefont
  {Comes}}, \bibinfo {author} {\bibfnamefont {M.}~\bibnamefont {Lambert}}, \
  and\ \bibinfo {author} {\bibfnamefont {A.}~\bibnamefont {Guinier}},\
  }\href@noop {} {\bibfield  {journal} {\bibinfo  {journal} {Solid State
  Commun.}\ }\textbf {\bibinfo {volume} {6}},\ \bibinfo {pages} {715} (\bibinfo
  {year} {1968})}\BibitemShut {NoStop}%
\bibitem [{\citenamefont {Ravel}\ \emph {et~al.}(1998)\citenamefont {Ravel},
  \citenamefont {Stern}, \citenamefont {Vedrinskii},\ and\ \citenamefont
  {Kraizman}}]{Ravel:1998vl}%
  \BibitemOpen
  \bibfield  {author} {\bibinfo {author} {\bibfnamefont {B.}~\bibnamefont
  {Ravel}}, \bibinfo {author} {\bibfnamefont {E.~A.}\ \bibnamefont {Stern}},
  \bibinfo {author} {\bibfnamefont {R.~I.}\ \bibnamefont {Vedrinskii}}, \ and\
  \bibinfo {author} {\bibfnamefont {V.}~\bibnamefont {Kraizman}},\ }\href@noop
  {} {\bibfield  {journal} {\bibinfo  {journal} {Ferroelectrics}\ }\textbf
  {\bibinfo {volume} {206}},\ \bibinfo {pages} {407} (\bibinfo {year}
  {1998})}\BibitemShut {NoStop}%
\bibitem [{\citenamefont {Qi}\ \emph {et~al.}(2016)\citenamefont {Qi},
  \citenamefont {Liu}, \citenamefont {Grinberg},\ and\ \citenamefont
  {Rappe}}]{Qi:2016}%
  \BibitemOpen
  \bibfield  {author} {\bibinfo {author} {\bibfnamefont {Y.}~\bibnamefont
  {Qi}}, \bibinfo {author} {\bibfnamefont {S.}~\bibnamefont {Liu}}, \bibinfo
  {author} {\bibfnamefont {I.}~\bibnamefont {Grinberg}}, \ and\ \bibinfo
  {author} {\bibfnamefont {A.~M.}\ \bibnamefont {Rappe}},\ }\href@noop {}
  {\bibfield  {journal} {\bibinfo  {journal} {ArXiv}\ ,\ \bibinfo {pages}
  {1601.02031v1}} (\bibinfo {year} {2016})}\BibitemShut {NoStop}%
\bibitem [{\citenamefont {Davis}\ \emph {et~al.}(2007)\citenamefont {Davis},
  \citenamefont {Budimir}, \citenamefont {Damjanovic},\ and\ \citenamefont
  {Setter}}]{Davis:2007gy}%
  \BibitemOpen
  \bibfield  {author} {\bibinfo {author} {\bibfnamefont {M.}~\bibnamefont
  {Davis}}, \bibinfo {author} {\bibfnamefont {M.}~\bibnamefont {Budimir}},
  \bibinfo {author} {\bibfnamefont {D.}~\bibnamefont {Damjanovic}}, \ and\
  \bibinfo {author} {\bibfnamefont {N.}~\bibnamefont {Setter}},\ }\href@noop {}
  {\bibfield  {journal} {\bibinfo  {journal} {J. Appl. Phys.}\ }\textbf
  {\bibinfo {volume} {101}},\ \bibinfo {pages} {054112} (\bibinfo {year}
  {2007})}\BibitemShut {NoStop}%
\bibitem [{\citenamefont {Hoshina}\ \emph {et~al.}(2014)\citenamefont
  {Hoshina}, \citenamefont {Kanehara}, \citenamefont {Takeda},\ and\
  \citenamefont {Tsurumi}}]{Hoshina2014}%
  \BibitemOpen
  \bibfield  {author} {\bibinfo {author} {\bibfnamefont {T.}~\bibnamefont
  {Hoshina}}, \bibinfo {author} {\bibfnamefont {K.}~\bibnamefont {Kanehara}},
  \bibinfo {author} {\bibfnamefont {H.}~\bibnamefont {Takeda}}, \ and\ \bibinfo
  {author} {\bibfnamefont {T.}~\bibnamefont {Tsurumi}},\ }\href@noop {}
  {\bibfield  {journal} {\bibinfo  {journal} {Jpn. J. Appl. Phys.}\ }\textbf
  {\bibinfo {volume} {53}},\ \bibinfo {pages} {09PD03} (\bibinfo {year}
  {2014})}\BibitemShut {NoStop}%
\bibitem [{\citenamefont {Misra}\ \emph {et~al.}(2004)\citenamefont {Misra},
  \citenamefont {Kotani}, \citenamefont {Kiwa}, \citenamefont {Kawayama},
  \citenamefont {Murakami},\ and\ \citenamefont {Tonouchi}}]{Misra2004}%
  \BibitemOpen
  \bibfield  {author} {\bibinfo {author} {\bibfnamefont {M.}~\bibnamefont
  {Misra}}, \bibinfo {author} {\bibfnamefont {K.}~\bibnamefont {Kotani}},
  \bibinfo {author} {\bibfnamefont {T.}~\bibnamefont {Kiwa}}, \bibinfo {author}
  {\bibfnamefont {I.}~\bibnamefont {Kawayama}}, \bibinfo {author}
  {\bibfnamefont {H.}~\bibnamefont {Murakami}}, \ and\ \bibinfo {author}
  {\bibfnamefont {M.}~\bibnamefont {Tonouchi}},\ }\href@noop {} {\bibfield
  {journal} {\bibinfo  {journal} {Appl. Surf. Sci.}\ }\textbf {\bibinfo
  {volume} {237}},\ \bibinfo {pages} {421} (\bibinfo {year}
  {2004})}\BibitemShut {NoStop}%
\bibitem [{\citenamefont {Hirori}\ \emph {et~al.}(2011)\citenamefont {Hirori},
  \citenamefont {Shinokita}, \citenamefont {Shirai}, \citenamefont {Tani},
  \citenamefont {Kadoya},\ and\ \citenamefont {Tanaka}}]{Hirori2011}%
  \BibitemOpen
  \bibfield  {author} {\bibinfo {author} {\bibfnamefont {H.}~\bibnamefont
  {Hirori}}, \bibinfo {author} {\bibfnamefont {K.}~\bibnamefont {Shinokita}},
  \bibinfo {author} {\bibfnamefont {M.}~\bibnamefont {Shirai}}, \bibinfo
  {author} {\bibfnamefont {S.}~\bibnamefont {Tani}}, \bibinfo {author}
  {\bibfnamefont {Y.}~\bibnamefont {Kadoya}}, \ and\ \bibinfo {author}
  {\bibfnamefont {K.}~\bibnamefont {Tanaka}},\ }\href@noop {} {\bibfield
  {journal} {\bibinfo  {journal} {Nat. Commun.}\ }\textbf {\bibinfo {volume}
  {2}},\ \bibinfo {pages} {594} (\bibinfo {year} {2011})}\BibitemShut {NoStop}%
\bibitem [{\citenamefont {Larson}\ and\ \citenamefont
  {Barhorst}(1980)}]{Larson1980}%
  \BibitemOpen
  \bibfield  {author} {\bibinfo {author} {\bibfnamefont {B.~C.}\ \bibnamefont
  {Larson}}\ and\ \bibinfo {author} {\bibfnamefont {J.~F.}\ \bibnamefont
  {Barhorst}},\ }\href@noop {} {\bibfield  {journal} {\bibinfo  {journal} {J.
  Appl. Phys.}\ }\textbf {\bibinfo {volume} {51}},\ \bibinfo {pages} {3181}
  (\bibinfo {year} {1980})}\BibitemShut {NoStop}%
\bibitem [{\citenamefont {Wie}\ \emph {et~al.}(1986)\citenamefont {Wie},
  \citenamefont {Tombrello},\ and\ \citenamefont {Vreeland}}]{Wie1986}%
  \BibitemOpen
  \bibfield  {author} {\bibinfo {author} {\bibfnamefont {C.~R.}\ \bibnamefont
  {Wie}}, \bibinfo {author} {\bibfnamefont {T.~A.}\ \bibnamefont {Tombrello}},
  \ and\ \bibinfo {author} {\bibfnamefont {T.}~\bibnamefont {Vreeland}},\
  }\href@noop {} {\bibfield  {journal} {\bibinfo  {journal} {J. Appl. Phys.}\
  }\textbf {\bibinfo {volume} {59}},\ \bibinfo {pages} {3743} (\bibinfo {year}
  {1986})}\BibitemShut {NoStop}%
\bibitem [{\citenamefont {Schick}\ \emph {et~al.}(2014)\citenamefont {Schick},
  \citenamefont {Herzog}, \citenamefont {Bojahr}, \citenamefont {Leitenberger},
  \citenamefont {Hertwig}, \citenamefont {Shayduk},\ and\ \citenamefont
  {Bargheer}}]{Schick:2014ie}%
  \BibitemOpen
  \bibfield  {author} {\bibinfo {author} {\bibfnamefont {D.}~\bibnamefont
  {Schick}}, \bibinfo {author} {\bibfnamefont {M.}~\bibnamefont {Herzog}},
  \bibinfo {author} {\bibfnamefont {A.}~\bibnamefont {Bojahr}}, \bibinfo
  {author} {\bibfnamefont {W.}~\bibnamefont {Leitenberger}}, \bibinfo {author}
  {\bibfnamefont {A.}~\bibnamefont {Hertwig}}, \bibinfo {author} {\bibfnamefont
  {R.}~\bibnamefont {Shayduk}}, \ and\ \bibinfo {author} {\bibfnamefont
  {M.}~\bibnamefont {Bargheer}},\ }\href@noop {} {\bibfield  {journal}
  {\bibinfo  {journal} {Struct. Dyn.}\ }\textbf {\bibinfo {volume} {1}},\
  \bibinfo {pages} {064501} (\bibinfo {year} {2014})}\BibitemShut {NoStop}%
\end{thebibliography}
\end{document}